\documentclass{article}
\usepackage{spconf,amsmath,graphicx, url, color}
\usepackage{hyperref}
\usepackage[table,xcdraw]{xcolor}
\usepackage{multirow}
\usepackage[normalem]{ulem}
\providecommand{\yz}[1]{\textcolor{black}{{#1}}}

\providecommand{\nz}[1]{\textcolor{black}{{#1}}}

\providecommand{\zd}[1]{\textcolor{black}{{#1}}}

\renewcommand{\sout}[1]{}

\title{SingFake: Singing Voice Deepfake Detection}
\name{Yongyi Zang*, You Zhang*, Mojtaba Heydari, Zhiyao Duan\thanks{* Equal contribution. This work is supported in part by a New York State Center of Excellence in Data Science award, \nz{National Institute of Justice (NIJ) Graduate Research Fellowship Award 15PNIJ-23-GG-01933-RESS}, National Science Foundation (NSF) grants 1846184, \zd{2222129}, and synergistic activities funded by NSF grant DGE-1922591.}}
\address{Department of Electrical and Computer Engineering, University of Rochester, Rochester, NY, USA}

\begin{document}
\ninept
\maketitle
\begin{abstract}
The rise of singing voice synthesis presents critical challenges to artists and industry stakeholders over unauthorized voice usage. Unlike synthesized speech, synthesized singing voices are typically released in songs containing strong background music that may hide synthesis artifacts. Additionally, singing voices present different acoustic and linguistic characteristics from speech utterances. These unique properties make singing voice deepfake detection a relevant but significantly different problem from synthetic speech detection. In this work, we propose the singing voice deepfake detection task. We first present SingFake, the first curated in-the-wild dataset consisting of 28.93 hours of bonafide and 29.40 hours of deepfake song clips in five languages from 40 singers. We provide a train/validation/test split where the test sets include various scenarios. We then use SingFake to evaluate four state-of-the-art speech countermeasure systems trained on speech utterances. We find these systems lag significantly behind their performance on speech test data. When trained on SingFake, either using separated vocal tracks or song mixtures, these systems show substantial improvement. However, our evaluations also identify challenges associated with unseen singers, communication codecs, languages, and musical contexts, calling for dedicated research into singing voice deepfake detection. The SingFake dataset and related resources are available\footnote{\url{https://www.singfake.org/}}.

\end{abstract}
\begin{keywords}
singing voice deepfake detection, anti-spoofing, dataset, singing voice separation
\end{keywords}
\section{Introduction}
\label{sec:intro}
\begin{quote}
\textit{``I mean really, how do you fight with someone who is putting out new albums in the time span of minutes.''} 

\hfill \textit{--- Stefanie Sun~\cite{makemusicsg2023}}
\end{quote}

Quoted from a prominent Singaporean singer, this remark underscores a rapidly emerging challenge in the modern music industry and cultural landscape: the surge of AI-generated singing voices. With the development of singing voice synthesis techniques, AI-generated singing voices sound increasingly natural, align well with the music scores, and can clone any singer's voice with a small amount of training data. Such techniques have been made more accessible with open-source singing voice synthesis and voice conversion projects, such as VISinger~\cite{zhang2022visinger} and DiffSinger~\cite{liu2022diffsinger}, raising concerns for artists, record labels, and publishing houses. For example, unauthorized synthetic productions mimicking a singer could potentially undermine the singer's commercial value, leading to potential copyright and licensing disputes. The ever-increasing societal apprehensions accentuate the urgency for developing methods to accurately detect deepfake singing voices.

As singing voice is a type of human vocalization, it is intuitive to explore solutions from an analogous research domain\yz{ - }speech deepfake detection, often referred to as voice spoofing countermeasures (CM). 
Existing research has been investigating different methods to discern speech spoofing attacks from bonafide human speech. Significant progress has been made in recent years. Contemporary state-of-the-art systems have showcased commendable performance, with some~\cite{jung2022aasist, xue2023learning, ding2023samo} achieving Equal Error Rate (EER) below 1\% on ASVspoof2019~\cite{wang2020asvspoof} test partitions. However, CM systems still suffer from generalization issues to unseen attacks and diverse acoustic environments, having shown strong degradation when evaluated on in-the-wild data~\cite{muller22_interspeech, liu2023asvspoof}. 


Singing voice deepfake detection, on the other hand, poses a distinct set of challenges not presented in speech. 
First, 
singing voices typically follow a specific melody or rhythmic pattern, which significantly affects the pitch and duration of different phonemes.
Second, singing voices have more artistic voicing traits and a wider range of timbre compared to speech, and \yz{are prone to influence by musical context.}
Lastly, singing voices often undergo extensive \yz{editing,} digital signal processing and are mixed with musical instrumental accompaniments.
Recognizing these unique \zd{properties of singing voices}, we question whether countermeasures developed for speech can be directly applied to singing voice spoof detection.


\begin{table*}[]
\caption{SingFake statistics for each split. 
}
\label{tab: singfake stats}
\centering
\begin{tabular}{c|lclc}
\hline \hline
Splits    & Description     & \# Singers                       & Languages (Sorted by percentages in the splits)       & \# Clips (Real / Fake)    \\ 
\hline
Train      & Training set    &      12                &  Mandarin,  Cantonese, Japanese, English,  Others &       5251 / 4519           \\ 
Val      & Validation set (unseen singers)     &     4                 &         Mandarin,  Cantonese, English, Spanish, Japanese                   &                  1089 / 543             \\ 
T01      & Test set for seen singer Stefanie Sun   &     1                 &     Mandarin,  Cantonese, Japanese, English, Others                       &               370 / 1208                       \\ 
T02  & Test set for unseen singers   &  6                &                Cantonese, Mandarin, Japanese            &                          1685 / 1006           \\ 
T03   &  T02 over 4 communication codecs &   6                 &                  Cantonese, Mandarin, Japanese          &                      6740 / 4024             \\ 
T04   & Test set for Persian musical context  &     17                 &       Persian, English                  &                   353 / 166                \\ 
\hline \hline
\end{tabular}
\end{table*}

In this paper, we propose the Singing Voice Deepfake Detection (SVDD) task. As \yz{a} first step, we curate the first in-the-wild dataset named SingFake to support this task. The SingFake dataset contains 28.93 hours of bonafide and 29.40 hours of deepfake song clips gathered from popular user-generated content platforms. Spanning five languages, we collect clips from 40 distinct singers and their AI counterparts. Additionally, we use a source separation model (Demucs~\cite{rouard2022hybrid}) to extract singing vocals from song mixtures, allowing us to examine the effects of singing vocals and song mixtures for SVDD systems separately.
We also provide a train/validation/test split, where the test set contains a diverse set of scenarios, including unseen singers, languages, communication codecs, and musical contexts.
With SingFake, we evaluate four types of leading speech countermeasure systems. We first use their models pretrained on speech utterances, and test them on the test split of SingFake. Results show a notable performance degradation compared to their performance on the ASVspoof2019 benchmark, on both song mixtures and separated vocals. We then retrain these systems on the training split of SingFake in two conditions, on separated vocal tracks and song mixtures, and test them on the test split. Results show significant improvement over the models trained on speech data. More detailed analyses of the results reveal challenges associated with unseen singers, communication codecs, languages, and musical contexts, underscoring the need for more focused research on crafting robust singing voice deepfake detection systems. 

During the process of writing this manuscript, we discovered another very recent study~\cite{xie2023fsd} that assesses the performance of speech CMs on clean singing voices and mixtures of them with instrumental music for Chinese songs. 
The study offers insight into speech CMs' capacity to learn deepfake cues under controlled ideal conditions, while our work focuses on more challenging in-the-wild scenarios. 




\section{SingFake Dataset Curation}
\subsection{Data collection and annotation}
We source deepfake singing samples from popular user-generated content websites where users upload both bonafide and deepfake samples of singing.
For every deepfake song sample, we manually annotate its metadata: AI singer, language, website, and label it as a ``deepfake'' song. The deepfake generation models, if disclosed, are also annotated.
We then collect bonafide samples from the corresponding real singer and annotate them as ``bonafide''. \nz{For deepfake samples, we \zd{identify} many AI-generated performances of songs in languages not sung by the original artists. Also, as the AI singers mimic celebrities, we \zd{verify} during annotation that those celebrities do not actually do the performances. } \yz{These steps \zd{ensure} the accuracy of our annotations.}
During the annotation process, we observe that most people use SoftVC-VITS\footnote{\url{https://github.com/svc-develop-team/so-vits-svc}} with different versions. The same uploader usually uses the same model to generate AI singer(s). 
As manual annotation is a tedious and error-prone process, to ensure the accuracy of metadata labels and correct potential inaccuracies, we employ GPT4~\cite{openai2023gpt} to verify the annotations against song titles and descriptions. We then manually \zd{review} any discrepancy found by GPT4. 
This process yields our SingFake dataset, capturing realistic deepfake nuances prevalent in online communities. SingFake contains 40 genuine singer and their corresponding AI singers, with 634 bonafide songs and 671 deepfake songs across multiple languages.
\yz{The SingFake dataset has been open-sourced under the CC BY-NC 4.0 license\footnote{\url{https://creativecommons.org/licenses/by-nc/4.0/deed}}.}



\subsection{Dataset splits}


Our primary guideline was to ensure that the singers were distinct across each section; therefore, we \zd{partition} the data into train/validation/test splits. Also, to form a comprehensive evaluation on the robustness of SVDD systems, we employ test subsets T01-T04 with increasing difficulties. Notably, there \zd{are} many samples from the singer "Stefanie Sun", so we set aside a portion of them, creating 
the T01 testing scenario to evaluate the system's performance on a seen-in-training singer. T02 testing set contains six unseen-in-training singers, while the T03 testing condition simulates the effect of \yz{media compression codecs} by encoding T02 through MP3 128 Kbps, AAC 64 Kbps, OPUS 64 Kbps, and Vorbis 64 Kbps. \nz{All song clips in T02 are passed through all 4 codecs to form T03.} The Persian singers \zd{stand} out as they contain mostly persian language while showing different musical styles. To investigate the effects of potential disparities in language and musical style, we allocate the Persian singers to a separate T04 test set. As T04 is collected from different platforms from other testing conditions, we believe T04 also contains unseen codecs. 

\begin{figure}[]
\centering
\includegraphics[width=0.43\textwidth]{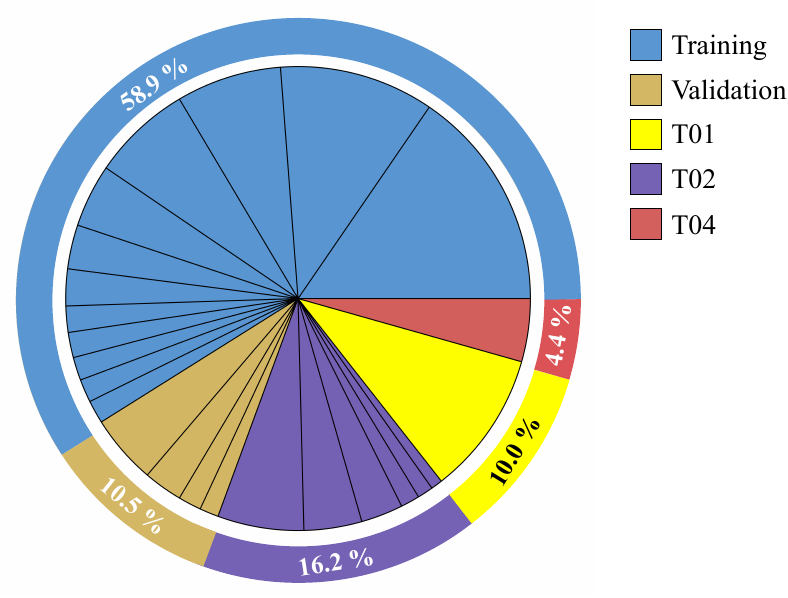}
\caption{SingFake dataset partition. Each color represents a subset, and each slice denotes an AI singer. T03 is excluded here since it contains the same song clips as T02 but is repeated 4 times through 4 different codecs.
}
\label{fig:pie}
\end{figure}

The rest of the dataset is split between training and validation, maintaining a rough song ratio of 6:1:1:2 between training, validation, T01, and T02 subsets.
Together, these subsets offer a comprehensive evaluation of the
singing voice deepfake detection systems. The final partitioning is illustrated in different colors in Figure~\ref{fig:pie}.


\subsection{Data processing}
Much of the data we collected comprises pop songs, which typically feature 
\yz{both instrumental sections, such as Intro, Outro and Interlude; and sections containing vocals, such as Verse and Chorus.}
For the purpose of singing voice deepfake detection, our primary interest lies in segments containing vocals, as only the singing voices are synthesized, rendering pure instrumental sections irrelevant for the SVDD task. Consequently, we narrowed our focus to regions with active singing, treating each as a distinct song clip. 

To extract active \yz{singing} regions, we first employed the state-of-the-art music source separation model, Demucs~\cite{rouard2022hybrid}\footnote{\url{https://github.com/facebookresearch/demucs}}, 
which is built on a U-Net convolutional architecture\yz{ with an attention-based bottleneck.}
This model is adept at separating vocals from musical accompaniments; the particular checkpoint we selected was trained on the MusDB~\cite{MUSDB18} dataset with extra training data, and secured the 2nd position on track B of the MDX challenge~\cite{mitsufuji2021music}. We use the separated vocals from Demucs as the source of separated song clips.

Next, the separated vocals are processed through the Voice Activity Detection (VAD) pipeline from PyAnnote \cite{pyannote}, which provides us with the timecodes for segmentation. These timecodes are subsequently used to segment both mixtures and vocals into individual song clips. 
\yz{All clips are resampled to 16 kHz during training and inference.}
For those songs originally in stereo, we maintained the stereo quality, but chose a random channel for each clip during training. 
The average length for clips in the dataset is 13.75 seconds.




The final statistics for all subsets including the splits at clip-level are shown in Table~\ref{tab: singfake stats}. We open-source the datasheet including original user-uploaded media links and our metadata annotations, dataset split generation and data processing code\footnote{\url{https://github.com/yongyizang/SingFake}}.


\section{Experiments}
In this section, we first evaluate existing speech spoofing countermeasure systems using SingFake. Subsequently, we retrain these systems from scratch using the SingFake training set and assess their performance across various test scenarios with our dataset splits.
\subsection{Experimental setup}
We construct four state-of-the-art systems that have demonstrated remarkable performance on speech datasets, representing different levels of input feature abstraction. This allows us to assess these features on both speech and singing spoof detection tasks.

\textbf{Model architectures}: \textbf{AASIST}~\cite{jung2022aasist} uses raw waveform as feature, leverages graph neural networks and incorporates spectro-temporal attention.
\textbf{Spectrogram+ResNet} uses a linear spectrogram extracted with 512-point FFT, with a hop size of 10 ms. We feed the extracted spectrogram into the ResNet18~\cite{he2016deep} architecture.
\textbf{LFCC+ResNet}~\cite{zhang2021one} uses Linear-Frequency Cepstral Coefficients (LFCC) as speech features, then feeds the LFCC into the ResNet18 model. The 60-dim LFCCs are extracted from each frame of the utterances, with frame length set to 20ms and hop size 10ms. 
\textbf{Wav2vec2+AASIST}~\cite{tak22_odyssey} is a model leveraging Wav2Vec2~\cite{baevski2020wav2vec}, a self-supervised front-end trained on large-scale external speech datasets. Note that we removed the RawBoost data augmentation module from the original paper~\cite{tak22_odyssey} for fair comparisons between methods, since no other method has such augmentation.

\textbf{Evaluation metric}:
Each system produces a score for each utterance, indicating the confidence that the given utterance is bonafide. 
The Equal Error Rate (EER) is determined by setting a threshold on the produced scores where the false acceptance rate matches the false rejection rate. \yz{EER is widely used as an indicator for biometric verification systems' performance}, and we think it is a good metric for SVDD as well.


\subsection{Speech CM heavily degrades on SVDD task}
We train and validate all speech CM systems on the speech dataset ASVspoof 2019 logical access (LA)~\cite{wang2020asvspoof} for 100 epochs. The model checkpoint with the best validation performance is selected for evaluation. 
We use the same train/dev/eval splits as ASVspoof 2019 LA. 
To form batches, we use 4 seconds of audio, \nz{following~\cite{jung2022aasist}}. We use repeat padding for shorter trials, and we randomly choose consecutive 4 seconds for longer trials. All of the CM systems achieve good performance on ASVspoof 2019 LA evaluation data, as shown in Table~\ref{tab:speech_res}. \nz{The results of AASIST and LFCC+ResNet are also comparable to the published ones~\cite{jung2022aasist, zhang2021one} while the wav2vec2+AASIST are not since we did not apply the data augmentation as they did.}

\begin{table}[]
\centering
\caption{Test results on speech and singing voice with CM systems trained on speech utterance from ASVspoof2019LA (EER (\%)).}
\label{tab:speech_res}
\begin{tabular}{cccccc}
\hline \hline
\multirow{2}{*}{\textbf{Method}} & {\textbf{ASVspoof2019}} &\multicolumn{2}{c}{\textbf{SingFake-T02}}   \\
                          &   \textbf{LA - Eval}  &  \textbf{Mixture}    &  \textbf{Vocals}      \\ \hline
AASIST                    & 0.83              & 58.12               & 37.91                \\

Spectrogram+ResNet               &   4.57            & 51.87               & 37.65                \\
LFCC+ResNet               & 2.41              & 45.12               & 54.88               \\
Wav2Vec2+AASIST              & 7.03                   & 56.75                     & 57.26                      \\
\hline \hline
\end{tabular}
\end{table}

\begin{table*}[]
\centering
\caption{Evaluation results for SVDD systems on all testing conditions in our SingFake dataset (EER (\%)) }
\label{tab:singing_test}
\setlength{\tabcolsep}{10pt} 
\begin{tabular}{cc| ccccc}
\hline \hline
\textbf{Method} &  \textbf{Setting} &  \textbf{Train} &  \textbf{T01} &  \textbf{T02} &  \textbf{T03} &  \textbf{T04} \\  \hline
\multirow{2}{*}{AASIST} & Mixture & \cellcolor[rgb]{0.98,0.98,0.98} \textcolor[rgb]{0.00,0.00,0.00}{4.10} & \cellcolor[rgb]{0.96,0.96,0.96} \textcolor[rgb]{0.00,0.00,0.00}{7.29} & \cellcolor[rgb]{0.93,0.93,0.93} \textcolor[rgb]{0.00,0.00,0.00}{11.54} & \cellcolor[rgb]{0.90,0.90,0.90} \textcolor[rgb]{0.00,0.00,0.00}{17.29} & \cellcolor[rgb]{0.77,0.77,0.77} \textcolor[rgb]{0.00,0.00,0.00}{\textbf{38.54}} \\
 & Vocals & \cellcolor[rgb]{0.98,0.98,0.98} \textcolor[rgb]{0.00,0.00,0.00}{3.39} & \cellcolor[rgb]{0.95,0.95,0.95} \textcolor[rgb]{0.00,0.00,0.00}{8.37} & \cellcolor[rgb]{0.94,0.94,0.94} \textcolor[rgb]{0.00,0.00,0.00}{10.65} & \cellcolor[rgb]{0.92,0.92,0.92} \textcolor[rgb]{0.00,0.00,0.00}{13.07} & \cellcolor[rgb]{0.74,0.74,0.74} \textcolor[rgb]{0.00,0.00,0.00}{43.94} \\

\multirow{2}{*}{Spectrogram+ResNet} & Mixture & \cellcolor[rgb]{0.97,0.97,0.97} \textcolor[rgb]{0.00,0.00,0.00}{4.97} & \cellcolor[rgb]{0.91,0.91,0.91} \textcolor[rgb]{0.00,0.00,0.00}{14.88} & \cellcolor[rgb]{0.86,0.86,0.86} \textcolor[rgb]{0.00,0.00,0.00}{22.59} & \cellcolor[rgb]{0.86,0.86,0.86} \textcolor[rgb]{0.00,0.00,0.00}{24.15} & \cellcolor[rgb]{0.71,0.71,0.71} \textcolor[rgb]{0.00,0.00,0.00}{48.76} \\
 & Vocals & \cellcolor[rgb]{0.97,0.97,0.97} \textcolor[rgb]{0.00,0.00,0.00}{5.31} & \cellcolor[rgb]{0.93,0.93,0.93} \textcolor[rgb]{0.00,0.00,0.00}{11.86} & \cellcolor[rgb]{0.88,0.88,0.88} \textcolor[rgb]{0.00,0.00,0.00}{19.69} & \cellcolor[rgb]{0.87,0.87,0.87} \textcolor[rgb]{0.00,0.00,0.00}{21.54} & \cellcolor[rgb]{0.74,0.74,0.74} \textcolor[rgb]{0.00,0.00,0.00}{43.94} \\
 \multirow{2}{*}{LFCC+ResNet} & Mixture & \cellcolor[rgb]{0.94,0.94,0.94} \textcolor[rgb]{0.00,0.00,0.00}{10.55} & \cellcolor[rgb]{0.87,0.87,0.87} \textcolor[rgb]{0.00,0.00,0.00}{21.35} & \cellcolor[rgb]{0.81,0.81,0.81} \textcolor[rgb]{0.00,0.00,0.00}{32.40} & \cellcolor[rgb]{0.81,0.81,0.81} \textcolor[rgb]{0.00,0.00,0.00}{31.85} & \cellcolor[rgb]{0.70,0.70,0.70} \textcolor[rgb]{0.00,0.00,0.00}{50.07} \\
 & Vocals & \cellcolor[rgb]{0.98,0.98,0.98} \textcolor[rgb]{0.00,0.00,0.00}{2.90} & \cellcolor[rgb]{0.90,0.90,0.90} \textcolor[rgb]{0.00,0.00,0.00}{15.88} & \cellcolor[rgb]{0.86,0.86,0.86} \textcolor[rgb]{0.00,0.00,0.00}{22.56} & \cellcolor[rgb]{0.86,0.86,0.86} \textcolor[rgb]{0.00,0.00,0.00}{23.62} &  \cellcolor[rgb]{0.76,0.76,0.76} \textcolor[rgb]{0.00,0.00,0.00}{39.27} \\
\multirow{2}{*}{Wav2Vec2+AASIST (Joint-finetune)} & Mixture &  \cellcolor[rgb]{0.99,0.99,0.99} \textcolor[rgb]{0.00,0.00,0.00}{\textbf{1.57}} &  \cellcolor[rgb]{0.97,0.97,0.97} \textcolor[rgb]{0.00,0.00,0.00}{\textbf{4.62}} &  \cellcolor[rgb]{0.95,0.95,0.95} \textcolor[rgb]{0.00,0.00,0.00}{\textbf{8.23}} & \cellcolor[rgb]{0.92,0.92,0.92} \textcolor[rgb]{0.00,0.00,0.00}{13.62} & \cellcolor[rgb]{0.74,0.74,0.74} \textcolor[rgb]{0.00,0.00,0.00}{42.77} \\
 & Vocals & \cellcolor[rgb]{0.99,0.99,0.99} \textcolor[rgb]{0.00,0.00,0.00}{1.70} & \cellcolor[rgb]{0.97,0.97,0.97} \textcolor[rgb]{0.00,0.00,0.00}{5.39} & \cellcolor[rgb]{0.95,0.95,0.95} \textcolor[rgb]{0.00,0.00,0.00}{9.10} &   \cellcolor[rgb]{0.94,0.94,0.94} \textcolor[rgb]{0.00,0.00,0.00}{\textbf{10.03}} & \cellcolor[rgb]{0.75,0.75,0.75} \textcolor[rgb]{0.00,0.00,0.00}{42.19} \\
\hline \hline
\end{tabular}
\end{table*}

We then test them on the T02 condition of SingFake to evalute their performance on singing data. All systems show heavy degradation as shown in Table~\ref{tab:singing_test}. The EERs are near 50\% on song mixtures, indicating that the speech deepfake detection systems are not able to distinguish real singers and their corresponding AI singers in the existence of accompanying music. 
Interestingly, both spectrogram-based and raw-waveform-based systems achieved around 38\% EER on the separated singing vocals, much better than the results on song mixtures. This might be due to the fact that singing vocals are more similar to speech compared to song mixtures since there would be nearly no music accompaniment presented after separation. However, the LFCC and Wav2Vec2-based systems are still performing near 50\% EER, indicating that these speech features tend to overfit more to the speech data and cannot generalize to singing voices.

\subsection{Training on singing voices improves SVDD performance}
\label{ssec:singing_train}


To investigate whether training on our curated SingFake dataset improves singing voice deepfake detection (SVDD) performance, we trained models using \yz{either full song mixtures (labeled as `Mixture') or separated singing vocals (labeled as `Vocals')}. Training on mixtures provides raw information, while training on separated vocals reduces instrumental distraction but may introduce separation artifacts that mask deepfake cues.

As shown in Table \ref{tab:singing_test}, SVDD performance declined from the training set (all seen) to T01 (seen singers, unseen songs) to T02 (unseen singers, unseen songs), indicating increasing task difficulty. All systems achieved good training set performance, showing that SingFake is helpful in learning the SVDD task. We also observed that the LFCC+ResNet system achieved the lowest training set performance on mixtures and the second-best performance on separated vocals, suggesting that instrumental interference may heavily hurt the spectral envelope. However, the noticeable decline in T02 performance highlights the challenge of generalizing SVDD to new singers. T01 performance fell between the training set and T02, suggesting that deepfakes of seen singers are easier to detect in new songs than those of unseen singers.

Compared to CM systems trained on speech, those trained on SingFake have better performance in terms of EER on T02, suggesting that the systems trained on SingFake are better at detecting singing deepfakes. The systems trained on separated vocals in general achieve better performance than those trained on mixtures except Wav2Vec2+AASIST. This suggests that separated singing voices could highlight artifacts for detecting singing deepfakes. 

Our results indicate that the Wav2Vec2+AASIST model excels in learning directly from song mixtures, delivering the most superior performance and robustness among all tested systems, similar to results reported for other tasks~\cite{tak22_odyssey, heydari2022singing}


\subsection{SVDD systems show limited robustness to unseen scenarios}


While training set, T01 and T02 represents more and more out-of-distribution sets at singer/song clips level, T03 and T04 sets are designed to evaluate performance in two challenging real-world situations: unseen communication codecs and unseen languages/musical contexts. Significant performance degradation is observed and well-studied under varying transmission and telecommunication codecs for speech CM systems~\cite{zhang21ea_interspeech, liu2023asvspoof}. However, when testing our system on the T03 condition, the performance drop was not as much as anticipated. As social media platforms typically employ a diverse set of audio compression codecs to more efficiently stream and deliver user-uploaded content, we believe the SingFake data we collected already utilizes codecs. Thus, when training the SVDD system, the model inherently learns to form a more robust representation that generalizes well across lossy audio compression algorithms.

At the same time, we observe significant performance degradation across all SVDD systems on T04, which is noticeably more pronounced than on both T02 and T03. T04 and T03 vary by unseen language and musical context, hinting that challenges posed by these attributes are still prominent for SVDD systems.
\section{Discussions}

The ability of AI to synthesize highly realistic singing voices demonstrate major technological progress, and also understandably cause public distrust, sometimes prompting calls to ban such technologies entirely. However, stopping advancement is rarely the answer. We believe transparency around content origins is key for establishing public trust, and more research into SVDD systems can allow users to make informed decisions about synthesized content. In this section, we summarize our findings on the strengths and weaknesses of SVDD systems.

\textbf{Unseen communication codecs.} \yz{We observed robust performance of SVDD systems against unseen compression codecs, as shown by the T03 example. This robustness differs from what we have seen with speech countermeasures. We hypothesize that this robustness stems from the SVDD systems being exposed to various compression codecs during training. Unlike speech deepfakes, most singing deepfakes are created for entertainment and are frequently posted on social media platforms where compression codecs are commonly applied. This entertainment focus and social media presence presumably leads to singing deepfakes incorporating more real-world compression compared to speech deepfakes.}


\textbf{Interference from backing tracks.} SVDD systems need to work on mixtures containing vocals and instrumental tracks, where the prominence of the instrumental tracks can make it challenging to detect deepfake vocals, since they may mask deepfake artifacts and introduce new artifacts that cause the systems to fail. While using source separation might mitigate this problem, as discussed in Section~\ref{ssec:singing_train}, any less-than-perfect separation result could inadvertently introduce new artifacts or mask deepfake cues, which may then confound the deepfake detection algorithms. 
\yz{As an example, we found that both the Demucs model and PyAnnote VAD pipeline tend to classify string instruments as active voice.}
This misclassification may have contributed to performance degradation on T04, as Persian music is rich in string instrumentation. 
\yz{This vulnerability calls for} developing interference-resilient SVDD systems and identifying more robust representations for this task.

\textbf{Diverse musical genres.} Singing voices in different genre follows significantly different musical context, exhibiting vastly different patterns of pitch, timbre and rhythm. 
By manual inspection, we discovered that the T04 subset contains many songs with heavy Hip-Hop influence, while most of the songs in other sets are rock and ballads. We believe this also contributes to the performance degradation seen on T04, \yz{indicating that SVDD systems fail to generalize to unseen musical genres.}
Since music reflects diverse cultural backgrounds, varying musical genres are likely to present in real-world SVDD situations. 
\yz{This vulnerability calls for future research efforts to} disentangle musical genre effects from deepfake cues, enabling more genre-agnostic SVDD systems.
As SVDD systems advance, we anticipate them to help enhance confidence in AI technologies within the music industry, restoring trust that have been eroded by the rise of \yz{unauthorized} deepfakes.




\section{Conclusions}

In this paper, we proposed the Singing Voice Deepfake Detection (SVDD) task and presented the Singfake dataset, containing a substantial collection of in-the-wild bonafide and deepfake song clips in various languages and singers. We demonstrated that state-of-the-art speech CM systems trained on speech show strong degradation when evaluated on singing voice, while re-training on singing voice leads to substantial improvements, highlighting the necessity of specialized SVDD systems.
Additionally, we assessed the strengths and weaknesses associated with unseen singers, communication codecs, different languages and musical contexts, underscoring the need for robust SVDD systems. Through releasing the SingFake dataset and benchmarking systems on the SVDD task, we aim to catalyze more research focused on developing specialized techniques for detecting deepfakes in singing voices.

\bibliographystyle{IEEEbib}
\bibliography{strings,refs}

\end{document}